\documentclass[final,3p,times]{elsarticle}

\usepackage{graphicx} 
\usepackage{tabularx}
\usepackage{amsmath}  
\usepackage{amssymb}
\usepackage{multirow}
\usepackage{subfigure}
\usepackage{epstopdf} 
\usepackage{color}
\usepackage{enumerate}

\usepackage{amssymb,amsmath,bm}

\begin{document}

\begin{frontmatter}

\title{ Boundary Variation Diminishing (BVD) reconstruction: a new approach to improve Godunov scheme }

\author[ad2]{Ziyao Sun}
\author[ad2]{Satoshi Inaba}
\author[ad2]{Feng Xiao \corref{cor}}

\address[ad2]{Department of Mechanical Engineering, Tokyo Institute of Technology, \\ 4259 Nagatsuta Midori-ku, Yokohama, 226-8502, Japan.}
\cortext[cor]{Corresponding author: Dr. F. Xiao (Email: xiao.f.aa@m.titech.ac.jp)}

\begin{abstract}
This paper presents a new approach, so-called boundary variation diminishing (BVD),  for reconstructions that minimize the discontinuities (jumps) at cell interfaces in Godunov type schemes. It is motivated by the observation that diminishing the jump at the cell boundary might effectively reduce the dissipation in numerical flux. Different from the existing practices which seek high-order polynomials within mesh cells while assuming discontinuities being always at the cell interfaces, we proposed a new strategy that combines a high-order polynomial-based interpolation and a jump-like reconstruction that allows a discontinuity being partly represented within the mesh cell rather than at the interface. It is shown that new schemes of high fidelity for both continuous and discontinuous solutions can be devised by the BVD guideline with properly-chosen candidate reconstruction schemes. 
Excellent numerical results have been obtained for both scalar and Euler conservation laws with substantially improved solution quality in comparison with the existing methods. This work provides a  simple and accurate alternative with great practical significance to the current Godunov paradigm which overly pursues the smoothness within mesh cell under the questionable premiss that discontinuities only appear at cell interfaces.
\end{abstract}

\begin{keyword}
Finite volume method \sep  Godunov method \sep discontinuity \sep reconstruction \sep Boundary Variation Diminishing (BVD)  \sep WENO  \sep THINC \sep  Riemann solver \sep compressible flow \sep shock wave.
\end{keyword}

\end{frontmatter}

\section{Introduction}

The high-resolution shock capturing schemes, which are now well-accepted as the main-stream numerical approach to solve hyperbolic conservation laws including the Euler equations for compressible gas dynamics, trace back to Godunov's scheme\cite{godunov59}. The original Godunov scheme is a conservative finite-volume method (FVM)  with piece-wise constant reconstruction, and the fluxes at cell boundaries are computed by solving the exact Riemann problem given the piece-wise constant physical fields in the two neighboring cells which usually result in a jump at cell boundaries. In general, Godunov schemes consist of two essential steps in solution procedure, i.e. (I) reconstruct the physical fields to find the values at the left- and right- sides of cell boundaries, and (II) evaluate the numerical fluxes at cell boundaries that are needed in the FVM formulation to update the cell-integrated values for next time step. 

For step (II), interested readers are referred to \cite{toro2009} for a monographic review. The main interest of this paper is limited to step (I), i.e. reconstruction. Continuous efforts have been devoted in the past half century to develop the original Godunov scheme into a class of high-resolution conservative schemes.  Instead of the peace-wise constant approximation which leads to a first-order scheme, higher order polynomials have been used to improve the accuracy (convergence rate for smooth solution). In order to get around the so-called Godunov barrier for linear schemes which states that any monotonic linear scheme can be of only first order,  non-linear numerical dissipation has been introduced to high-order schemes to suppress spurious oscillations in the vicinity of discontinuities. The non-linear numerical dissipation has been devised in different forms seen in the literature as either flux limiters in flux-corrected transport (FCT) scheme\cite{boris1973,zalesak1979} and total variation diminishing (TVD) scheme \cite{harten1983,sweby1984} or slope limiters in monotone upstream-centered schemes for conservation law (MUSCL) scheme \cite{vleer77,vleer79}. The later gives a more straightforward interpretation of the polynomial-based reconstruction for high-order schemes, and shows a clear path to design high-resolution schemes using modified high-order polynomials. Representative schemes of this kind are piecewise parabolic method (PPM) method\cite{colella1984}, essentially non-oscillatory scheme(ENO) \cite{harten1987,shu1988,shu1988} and weighted essentially non-oscillatory (WENO)  scheme \cite{liu1994,jiang1996,shu2009}. WENO concept provides a general framework to develop high-resolution schemes that are able to reach the highest possible accuracy over a given mesh stencil for smooth solution, and effectively stablize the numerical solutions which include discontinuities. Successive works on WENO scheme are found in designing better smoothness indicators and nonlinear weights \cite{weno-m,weno-z,arshed2013,sz14,hu2015} to improve solution quality, as well as implementing the WENO concept to different discretization frameworks\cite{deng2000,pirozzoli2002,ren2003,liu2015}. 

Starting from the piecewise constant approximation in the original Godunov scheme, the development of  high-order shock capturing schemes has evidenced a history of efforts to reduce the difference between the reconstructed left- and right-side values for smooth solutions. This observation, however to our knowledge, has not been paid enough attention,  nor explored further as a prospective guideline to construct new schemes.  

In this work, we propose  a new guideline to design high-resolution schemes, called boundary variation diminishing (BVD) reconstruction. The BVD reconstruction shows consistent results with the observation gained for smooth solutions as mentioned above. More importantly, it also provides a principle to devise the reconstruction for  
discontinuity. We demonstrate in this paper the BVD reconstruction based on two existing schemes, i.e. the 5th-order WENO method\cite{weno-z} and the THINC (Tangent of Hyperbola for INterface Capturing) method\cite{xiao2005,xiao2011,shyue2014}. 

\section{Numerical algorithm}
We use a scalar conservation law in the following form  to present the BVD algorithm.
\begin{equation}\label{1D-scalar-transport-equation}
{{\partial u} \over {\partial t}}+{{\partial f} \over {\partial x}} =0,  
\end{equation}
where $ u(x) $ is the solution function and $ f(u) $ is the flux function.  For a hyperbolic equation, $\alpha=f^{'}(u)$ is a real number, the characteristic speed. 

We divide the computational domain into $N$ non-overlapping cell elements $I_i=[x_{i-{1\over2}},x_{i+{1\over2}}]$, $i=1,2,\cdots,N$. The mesh is assumed to be uniform across the computational domain for simplicity, $\Delta x=x_{i+\frac{1}{2}}-x_{i-\frac{1}{2}} $. 

Using an FVM framework, we define the volume-integrated average VIA of a function $ u(x) $ for cell $I_i$ as, 
\begin{equation}\label{VIA-moment}
\overline{u}_{i}=\frac{1}{\Delta x}\int_{x_{i-{1\over2}}}^{x_{i+{1\over2}}}u(x,t) dx.
\end{equation}

The VIA $ \bar{u}_{i} $ of each cell $I_i=[x_{i-{1\over2}},x_{i+{1\over2}}]$ can be updated by
\begin{equation}
\frac{d \bar{u}_i}{dt}=-\frac{1}{\Delta x}(\tilde{f}_{i+\frac{1}{2}}-\tilde{f}_{i-\frac{1}{2}}), 
\end{equation}
where the numerical fluxes at cell boundaries are computed by a Riemann solver, 
\begin{equation}\label{Numerical Flux}
\tilde{f}_{i+\frac{1}{2}}={f}^{\rm Riemann}_{i+\frac{1}{2}}(u_{i+\frac{1}{2}}^{L},u_{i+\frac{1}{2}}^{R}), 
\end{equation}
using the left-side value  $u^{L}$ and right-side value  $u^{R}$ obtained from the reconstructions over left- and right-biased stencils. In spite of different variants, the Riemann flux is essentially upwinding and can be thus written in a canonical form as, 
\begin{equation}\label{Lax-Friedrichs}
{f}^{\rm Riemann}_{i+\frac{1}{2}}(u_{i+\frac{1}{2}}^{L},u_{i+\frac{1}{2}}^{R})=\frac{1}{2}\left(f(u_{i+\frac{1}{2}}^{L})+f(u_{i+\frac{1}{2}}^{R})-|\tilde{\alpha}_{i+\frac{1}{2}}|(u_{i+\frac{1}{2}}^{R}-u_{i+\frac{1}{2}}^{L})\right),
\end{equation}
where $\tilde{\alpha}_{i+\frac{1}{2}}$ stands for a characteristic speed in a hyperbolic equation. The last term in \eqref{Lax-Friedrichs} can be interpreted as a numerical dissipation.

Our central task now is how to calculate the left-side value  $u_{i+\frac{1}{2}}^{L}$ and right-side value  $u_{i+\frac{1}{2}}^{R}$ for all cell boundaries $x_{i+\frac{1}{2}}$, $i=1,2,\cdots,N$.

\subsection{Boundary variation diminishing (BVD) reconstruction} \label{BVDr}

We propose a novel guideline for reconstruction that adaptively chooses proper interpolation functions so as to minimize the jump between the left- and right- side values, $u^L$ and $u^R$, at cell interface. We elaborate the algorithm of boundary variation diminishing (BVD) reconstruction for cell $I_i$ as follows.

\begin{enumerate}[i)]
\item Prepare two piece-wisely reconstructed interpolation functions, $\Phi^{<1>}_i(x)$ and $\Phi^{<2>}_{i}(x)$,  of solution function $u(x)$ for each cell from the VIAs $\bar{u}_i$ available over the computational domain. The existing reconstruction methods, such as the WENO and THINC described later, can be employed for this purpose. Without losing generality, we usually assume that $\Phi^{<1>}_i(x)$ is a higher-order polynomial-based interpolant, while  $\Phi^{<2>}_{i}(x)$ might be of low-order but has better monotonicity (a sigmoid function is preferred for step-like discontinuity, for example);  
\item Find $\Phi^{<p>}_i(x)$ and $\Phi^{<q>}_{i+1}(x)$ with $p$ and $q$ being either 1 or 2, so that the boundary variation (BV)
\begin{equation} \label{min_ip12}
BV (\Phi)_{i+\frac{1}{2}} =|\Phi^{<p>}_i(x_{i+\frac{1}{2}})-\Phi^{<q>}_{i+1}(x_{i+\frac{1}{2}})| 
\end{equation}
is minimized;
\item In case that a different choice for cell $i$ is resulted in when applying step ii) to the neighboring interface $x_{i-\frac{1}{2}}$, that is,  $\Phi^{<p>}_i(x)$ found to minimize 
\begin{equation}\label{min_im12}
BV (\Phi)_{i-\frac{1}{2}} =|\Phi^{<p>}_{i-1}(x_{i-\frac{1}{2}})-\Phi^{<q>}_{i}(x_{i-\frac{1}{2}})| 
\end{equation}
is different from that found to minimize \eqref{min_ip12}, we adopt the following criterion to uniquely determine the reconstruction function. 
\begin{equation}
\Phi^{<p>}_{i}(x)=
\left\{ \begin{array}{l}
\Phi^{<1>}_{i}(x),  \  {\rm if} \ \left(\Phi^{<p>}_i(x_{i+\frac{1}{2}})-\Phi^{<q>}_{i+1}(x_{i+\frac{1}{2}})\right)\left(\Phi^{<p>}_{i-1}(x_{i-\frac{1}{2}})-\Phi^{<q>}_{i}(x_{i-\frac{1}{2}}\right) < 0, \\ 
\Phi^{<2>}_{i}(x),  \  {\rm otherwise. }
\end{array}\right.
\end{equation}

\item  Compute the left-side value $ u^{L}_{i+\frac{1}{2}} $ and the right-side value $ u^{R}_{i-\frac{1}{2}} $ for each cell by 
\begin{equation}
u^{L}_{i+\frac{1}{2}}=\Phi^{<p>}_i(x_{i+\frac{1}{2}}) \ \ {\rm and} \ \  u^{R}_{i-\frac{1}{2}}=\Phi^{<p>}_{i}(x_{i-\frac{1}{2}}).
\end{equation}
 \end{enumerate}

\begin{description}
\item {Remark 1. } 
The BVD algorithm reduces the reconstructed jumps at cell  interfaces, and thus the numerical dissipation term $ND_{i+\frac{1}{2}}=|\tilde{\alpha}|_{i+\frac{1}{2}}(u_{i+\frac{1}{2}}^{R}-u_{i+\frac{1}{2}}^{L})$ in Riemann solvers, which can be expected to effectively improve the numerical solution. 
\item {Remark 2. } 
For smooth solution, the BVD reconstruction naturally realizes the highest possible interpolation because interpolants of higher order tend to find an interface value closer to the ``true" solution.  
\item {Remark 3. }   
For discontinuous solution, pursuing higher order polynomials does not necessarily lead to the reduction of the reconstructed boundary jumps. It might imply the limitation of the current practice to use high-order polynomials for reconstruction. 
\item  {Remark 4. }
By minimizing the jumps at cell boundaries, the BVD algorithm requires a reconstruction that is able to represent a jump within the cell where a discontinuity exists.  In this sense, the THINC reconstruction shown later is preferred to a polynomial.  
\item {Remark 5. }  The BVD reconstruction provides practical and effective guideline to construct high-fidelity schemes for resolving discontinuous solutions. As shown later, excellent numerical results can be obtained if the candidate reconstructions, $\Phi^{<1>}_i(x)$ and $\Phi^{<2>}_{i}(x)$,  are properly chosen. 
\item  {Remark 6. } Other  BV-equivalent quantities can be also used in the algorithm. For example, we have tested to minimize  the total boundary variation (TBV),  $TBV (\Phi)_i=|BV (\Phi)_{i-\frac{1}{2}}|+|BV (\Phi)_{i+\frac{1}{2}}|$  for cell $i$, which gives quite similar results but is more algorithmically complicated.  Being aware of that the numerical dissipation appears eventually in a numerical scheme as $ND_{i+\frac{1}{2}}-ND_{i-\frac{1}{2}}$, we see another future option to minimize $|ND_{i+\frac{1}{2}}-ND_{i-\frac{1}{2}}|$.    
\end{description}	

\subsection {Two candidate reconstruction schemes}

As shown in the BVD reconstruction algorithm, we need two building-block schemes to construct the candidate interpolation functions,  $\Phi^{<1>}_i(x)$ and $\Phi^{<2>}_{i}(x)$.  In order to enable the resulting scheme to accurately resolve both smooth and discontinuous solutions. We use the fifth-order WENO scheme and the THINC scheme for reconstructions, and give a brief description below.

\subsubsection{The 5th-order WENO reconstruction}\label{WENOZ reconstuction}

We choose the 5th-order WENO scheme \cite{weno-z} as $\Phi^{<1>}_i(x)$, and compute the cell interface values on an uniform grid  by  
\begin{equation}
\Phi^{<1>}_i(x_{i\pm\frac{1}{2}})=\omega_{0}u^{(0)}_{i\pm\frac{1}{2}}+\omega_{1}u^{(1)}_{i\pm\frac{1}{2}}+\omega_{2}u^{(2)}_{i\pm\frac{1}{2}},
\end{equation}
where $u^{(0)}$,  $u^{(1)}$ and  $u^{(2)}$ are computed from the third-order polynomials constructed separately over three slided 4-cell stencils. 

\begin{equation}
\begin{aligned}
			&\hspace{0pt}u^{(0)}_{i-\frac{1}{2}}=\frac{1}{3}\overline{u}_{i+2}-\frac{7}{6}\overline{u}_{i+1}+\frac{11}{6}\overline{u}_{i}, \ u^{(1)}_{i-\frac{1}{2}}=-\frac{1}{6}\overline{u}_{i+1}+\frac{5}{6}\overline{u}_{i}+\frac{1}{3}\overline{u}_{i-1}, \ u^{(2)}_{i-\frac{1}{2}}=\frac{1}{3}\overline{u}_{i}+\frac{5}{6}\overline{u}_{i-1}-\frac{1}{6}\overline{u}_{i-2}; \\
			&\hspace{0pt}u^{(0)}_{i+\frac{1}{2}}=\frac{1}{3}\overline{u}_{i-2}-\frac{7}{6}\overline{u}_{i-1}+\frac{11}{6}\overline{u}_{i}, \ u^{(1)}_{i+\frac{1}{2}}=-\frac{1}{6}\overline{u}_{i-1}+\frac{5}{6}\overline{u}_{i}+\frac{1}{3}\overline{u}_{i+1}, \ u^{(2)}_{i+\frac{1}{2}}=\frac{1}{3}\overline{u}_{i}+\frac{5}{6}\overline{u}_{i+1}-\frac{1}{6}\overline{u}_{i+2}.
\end{aligned}
\end{equation}	

The computation of the non-linear weights, $ \omega_{j}$, $j=0,1,2$, depends on the smoothness indicator  which measures the smoothness of solution function $u(x)$. We follow the WENO-Z scheme to evaluate the nonlinear weights, see \cite{weno-z} for more details.

\subsubsection{The THINC reconstruction}\label{THINC reconstuction}
Being a sigmoid function, hyperbolic tangent function is a differentiable and monotone function that fits well a step-like discontinuity. A class of VOF(volume of fluid) schemes, so-called THINC (Tangent of Hyperbola for INterface Capturing), has been devised to compute moving interfaces in multiphase flow simulations \cite{xiao2005,xiao2011,shyue2014} based on the hyperbolic tangent function. In the present work, we use THINC reconstruction as the second candidate interpolation function  $\Phi^{<2>}_i(x)$ in the BVD algorithm, 
\begin{equation}
\Phi^{<2>}_{i}(x)=u_{min}+\frac{u_{max}}{2}\left( 1+\gamma\tanh\left( \beta\left( \frac{x-x_{i-\frac{1}{2}}}{x_{i+\frac{1}{2}}-x_{i-\frac{1}{2}}}-\tilde{x}_{i}\right) \right) \right), 
\end{equation}
where  $ u_{min}=\min(\bar{u}_{i-1},\bar{u}_{i+1})$, $ u_{max}=\max(\bar{u}_{i-1},\bar{u}_{i+1})-u_{min}$ and $ \gamma={\rm sgn}(\bar{u}_{i+1}-\bar{u}_{i-1}) $.  Parameter $\beta $ is used to control the jump thickness. We use a  constant value of $ \beta =1.6$ for all numerical tests in this paper. The unknown $ \tilde{x}_{i}$, which represents the location of the jump center, is computed from the constraint condition, 
$\bar{u}_{i}=\frac{1}{\Delta x}\int_{x_{i-\frac{1}{2}}}^{x_{i+\frac{1}{2}}}\Phi^{<2>}_{i}(x)dx.$

Rather than showing explicitly the formula to calculate $\tilde{x}_{i}$,  we give the expressions for the reconstructed left- and right-end values of cell $ i $ by
\begin{equation}
\begin{aligned}
&\hspace{0pt}\Phi^{<2>}_{i}(x_{i-\frac{1}{2}})=u_{min}+\frac{u_{max}}{2}\left(1+\gamma\frac{\tanh(\beta)+A}{1+A\tanh(\beta)}\right),\\
&\hspace{0pt}\Phi^{<2>}_{i}(x_{i+\frac{1}{2}})=u_{min}+\frac{u_{max}}{2}(1+\gamma A),
\end{aligned}
\end{equation}
where $A=\frac{  B/\cosh(\beta)-1  }{\tanh(\beta)}, \ B=\exp\left(\gamma\beta\left( 2\frac{\bar{u}_{i}-\bar{u}_{min}+\epsilon}{u_{max}+\epsilon}-1\right) \right)$ and $\epsilon=10^{-20}$. It should be noted that the THINC reconstruction applies only to cells where $\left(\overline{u}_{i+1}-\overline{u}_{i}\right)\left(\overline{u}_{i}-\overline{u}_{i-1}\right)>0$ hold.

\begin{description}
\item {Remark 7. } 
We also experimented using piece-wise constant and MUSCL reconstructions for $\Phi^{<2>}_{i}(x)$, and find that the BVD algorithm results in a less oscillatory solution without noticeable increase in numerical dissipation in comparison with the WENO-only computation. Another combination with MUSCL for $\Phi^{<1>}_{i}(x)$ and THINC for $\Phi^{<2>}_{i}(x)$ gives significant improvement in resolving contact discontinuities in numerical solutions.  
\end{description}

We refer to the BVD algorithm with WENO and THINC reconstructions presented in this paper as BVD-WENO-THINC method. 
		 
\subsection{Some remarks on implementation}		 
We summarize some important points regarding our implementation of BVD in the present work. 
		 \begin{itemize}
		 	\item We use the WENO reconstruction for $\Phi^{<1>}_{i}(x)$ and THINC reconstruction for  $\Phi^{<2>}_{i}(x)$, and find the left-side and right-side values  $ u^{L}_{i+\frac{1}{2}} $ and $u^{R}_{i+\frac{1}{2}} $ for each cell interface following the BVD reconstruction procedure described in \ref{BVDr}. 
		 	\item For Euler equations, the minimization in BVD reconstruction is conducted in terms of  the primitive variables separately, i.e. density, velocity and pressure. We use  Roe's Riemann solver\cite{roe1981} to calculate the numerical fluxes. 
		 	\item The five stage fourth order SSP Runge-Kutta method \cite{spiteri2002new} is used for time marching with a CFL number of 0.4. 
		 \end{itemize}

\section{Numerical results}			
In this section, the numerical results for some benchmark tests are presented  to illustrate the performance of our new algorithm.  The detail set-ups of  numerical experiments  can be found in \cite{jiang1996,liu1994}. 
\subsection{Scalar Conservation law}
\subsubsection{Linear advection equation}
Firstly, we consider the following 1D linear advection equation,
\begin{equation}
	u_{t}+u_{x}=0.
\end{equation}

\textbf{Accuracy test for 1D advection equation.} An initially smooth profile defined by $ u(x,0)=\sin(\pi x),\ x\in[-1,1]$ is advected. The $ L_{1} $ and $ L_{\infty} $ errors of VIA after one period are measured under different grid resolutions. Shown in Table \ref{Adv-Accu}, the numerical errors of  BVD-WENO-THINC are identical to that of WENO-Z for all grid resolutions. It proves that the BVD-WENO-THINC scheme retrieves the $ 5th $ order WENO schemes for smooth solution.
\begin{table}[h]
			\centering
	\caption{Numerical errors and convergence rate for 1D advection equation, $ t=2.0 $. }
	\label{Adv-Accu}
	\begin{tabular}{ccccccccccc}
		\hline $ N$  & \multicolumn{4}{c}{BVD-WENO-THINC scheme}& & \multicolumn{3}{c}{WENO-Z scheme} \\ 
		\cline{2-5} \cline{7-10}&$  L_{1}$ error & Order & $ L_{\infty} $    & Order&  & $  L_{1}$ error  & Order &$ L_{\infty} $& Order\\ 
		$ 20$    & 2.14e-04   &      & 3.65e-06 &  & & 2.14e-04   &      & 3.65e-06   \\ 
		$ 40$    & 6.40e-06  &5.07 & 1.03e-05 & 5.10&& 6.40e-06  &5.07 & 1.03e-05 & 5.10\\ 
		$ 80$    & 2.00e-07  & 5.00 & 3.18e-07 & 5.02&& 2.00e-07  & 5.00 & 3.18e-07 & 5.02\\
		$ 160$   & 6.32e-09  & 4.99 & 9.96e-09 & 5.00&& 6.32e-09  & 4.99 & 9.96e-09 & 5.00\\  
		$ 320$   & 2.04e-10  & 4.00 & 3.20e-10 & 4.96&& 2.04e-10  & 4.00 & 3.20e-10 & 4.96\\ 
		\hline 
	\end{tabular}
\end{table}

\begin{figure}[hp]
	\centering
	\includegraphics[width=4in]{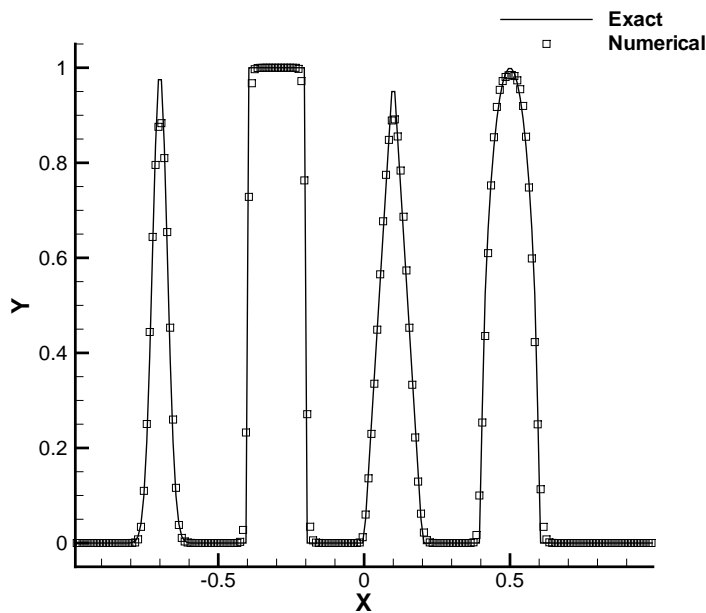}
	\caption{Numerical results of Jiang and Shu's test after one period ($ t=2.0 $) with $ 200 $ grid cells.}
	\label{Shu's test}
\end{figure}

\textbf{Jiang and Shu's test.} The initial condition of this test includes both discontinuities and smooth profile. We set the initial condition as
\begin{equation}
u(x,0)=
\left\lbrace 
\begin{aligned}
&\hspace{0pt}\frac{1}{6}\left( G(x,\beta,z-\delta)+G(x,\beta,z+\delta)+4G(x,\beta,z)\right),   \quad  -0.8\leqslant x\leqslant -0.6,\\
&\hspace{0pt}1, \qquad \qquad \qquad  \qquad \qquad \qquad \qquad \qquad \qquad  \ \ \ -0.4\leqslant x\leqslant -0.2,\\
&\hspace{0pt}1-\lvert 10(x-0.1) \rvert,   \qquad \qquad \qquad \qquad \qquad \qquad \quad \quad \  0.0\leqslant x\leqslant 0.2,\\
&\hspace{0pt}\frac{1}{6}\left( F(x,\alpha,a-\delta)+F(x,\alpha,a+\delta)+4F(x,\alpha,a)\right),   \quad \quad \ 0.4\leqslant x\leqslant 0.6,\\
&\hspace{2pt}0,  \qquad \qquad \qquad \qquad \qquad \qquad \qquad \qquad \qquad \qquad  \rm{otherwise}, \\
\end{aligned}
\right.
\end{equation} 
where the computation domain is $ [-1,1] $. The function $ F $ and $ G $ is defined by
\begin{equation}
G(x,\beta,z)=\exp\left( -\beta(x-z)^{2}\right) , F(x,\alpha,a)=\sqrt{\max(1-\alpha^{2}(x-a)^{2},0)}, 
\end{equation}
and the coefficients to determine the initial profile are given by
\begin{equation}
a=0.5,\ z=0.7,\ \delta=0.005, \ \alpha=10.0, \ \beta=\log 2/(36\delta^{2}).
\end{equation}
The numerical results are shown in Fig.~\ref{Shu's test}. It can be observed that our scheme can resolve the sharp discontinuity while keeping the smooth profile with good resolution.

\subsubsection{Inviscid Burgers equation}
We consider the 1D inviscid Burger's equation
\begin{equation}
u_{t}+\left( \dfrac{u^{2}}{2}\right)_{x}=0.
\end{equation}
Here we will give two examples for inviscid Burgers equation.

\begin{figure}[hp]
		\centering
		\includegraphics[width=8cm]{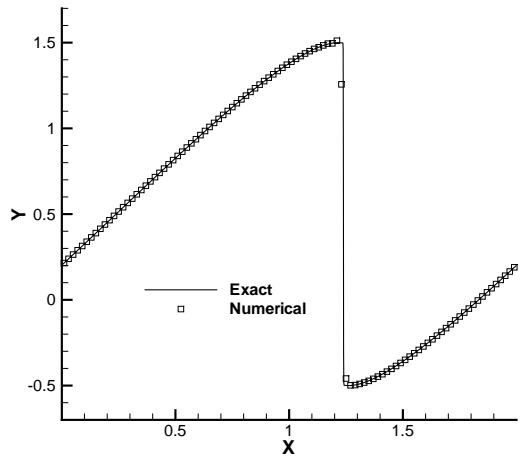} 
		\caption{Numerical results of Burgers equation at $ t=1.5/\pi $ on a $100  $ cell-mesh. }
		\label{1D Burger equation-1}
	\end{figure}

\textbf{Burgers equation with an initial  sine wave.} In this test, a sine wave is given as the initial profile, which will develop into a shock wave after $ t=1.5/\pi $ due to the convexity of the Burgers flux. Firstly, in order to measure the convergence rate of our scheme, we carried out the computation up to $ t=0.5/\pi $. The numerical errors are shown in Table \ref{Accuracy test Burger equation}, which demonstrate nearly the $ 5th  $ order accuracy.  We also show the numerical results computed on a 100-cell mesh at $ t=1.5/\pi $ in Fig~\ref{1D Burger equation-1}. We can see that the shock wave in this test can be resolved within only two cells.

	\begin{table}[hp]
		\centering
		\caption{Numerical errors and convergence rates for 1D Burgers equation at $ t=1/(2\pi). $} 
		\begin{tabular}{cccccc}
			\hline $ N$& $  L_{1}$ error & Order of Accuracy  & $ L_{\infty}$   error   & Order of Accuracy  & \\  
			\hline $ 20$        & 3.00e-04  &  & 2.07e-03 &  \\ 
			$ 40$       & 1.66e-05  & 4.18 & 2.15e-04 & 3.27 \\ 
			$ 80$       & 7.36e-07  & 4.49 & 9.94e-06 & 4.43 \\ 
			$ 160$       & 2.65e-08  & 4.79 & 3.56e-07 & 4.80 \\ 
			\hline 
		\end{tabular} 
		\label{Accuracy test Burger equation}
	\end{table}

\begin{figure}[hp]
	\centering
	\includegraphics[width=4.0in]{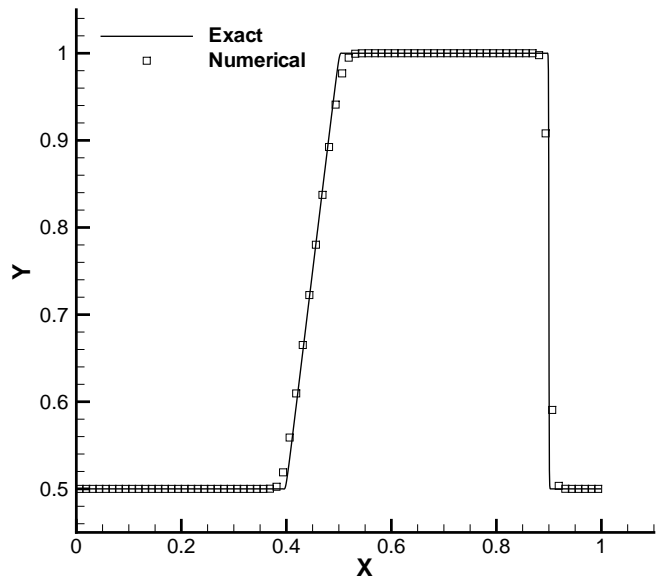} 
	\caption{Numerical results of Burgers equation with a shock and rarefaction at $ t=0.2 $ with $80$ cells. }
	\label{1D Burger equation-2}
\end{figure}

\textbf{Burgers equation with shock and rarefaction waves.} The initial condition of this benchmark test is given by
	\begin{equation}
	u(x,0)=
	\left\lbrace 
	\begin{aligned}
	&\hspace{0pt}1,  \quad 0.3 \leqslant x\leqslant 0.75,\\
	&\hspace{0pt}0.5, \quad \rm{otherwise},
	\end{aligned}
	\right.
	\end{equation}
Both shock and rarefaction wave are included in this test. The numerical results at $ t=0.2 $ are shown in Fig~\ref{1D Burger equation-2}. The BVD-WENO-THINC scheme can recover both the shock and expansion fan with good accuracy. 	

\subsubsection{Buckley Leverett equation}

\begin{figure}[hp]
	\centering
	\includegraphics[width=4.0in]{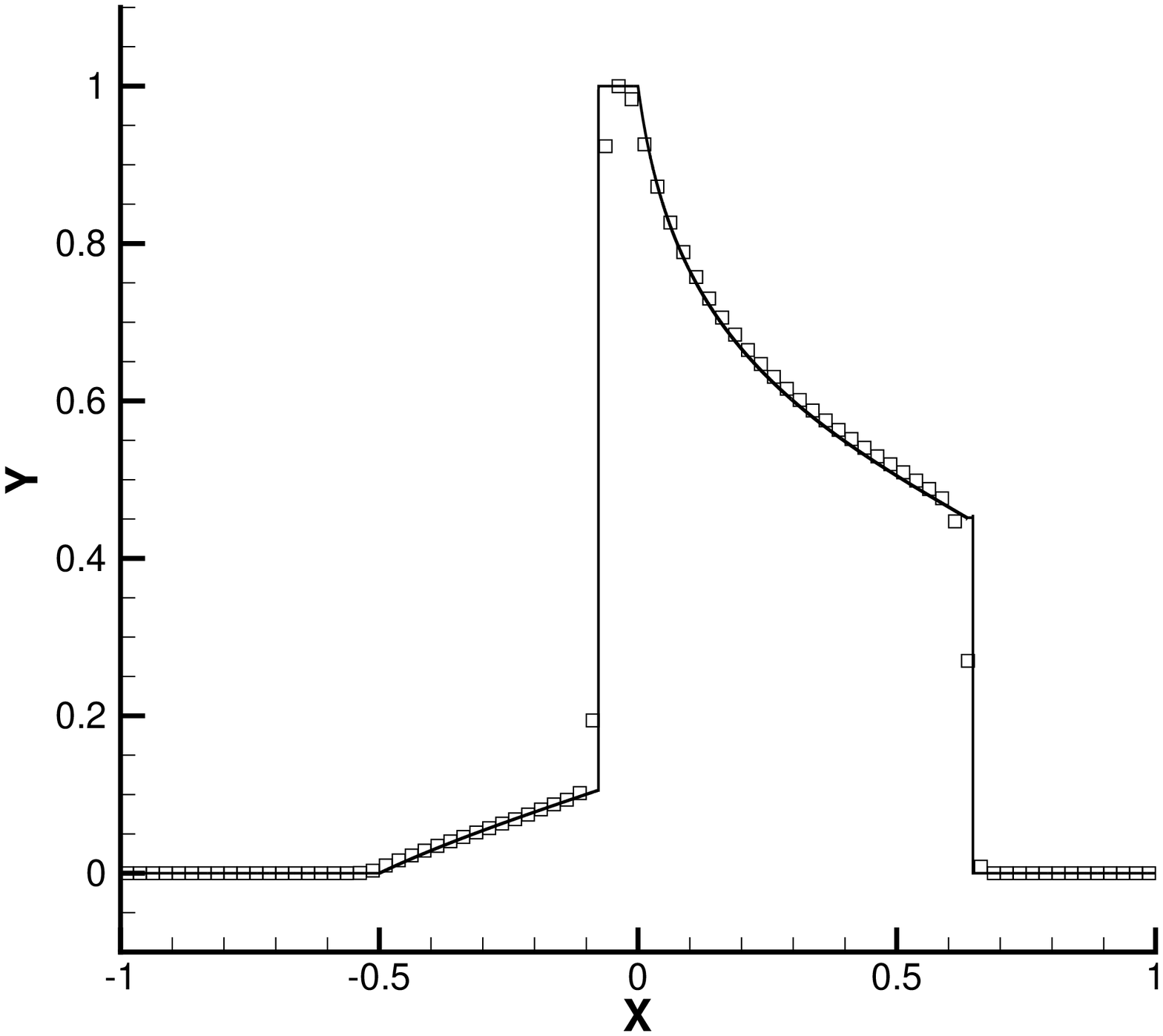} 
	\caption{Numerical solutions of Buckley Leverett problem at $ t=0.2 $ with $80  $ cells. }
	\label{1D BL-1}
\end{figure}

1D Buckley Leverett equation reads 
\begin{equation}
u_{t}+\left( \dfrac{4u^{2}}{4u^{2}+(1-u)^{2}}\right)_{x}=0.
\end{equation}
\textbf{Square pulse problem.}
The initial condition is set to be a square pulse, 
\begin{equation}
u(x,0)=
\left\lbrace 
\begin{aligned}
&\hspace{0pt}1,  \quad -\frac{1}{2}\leqslant x\leqslant 0,\\
&\hspace{0pt}0,\ \quad \rm{otherwise}.
\end{aligned}
\right.
\end{equation}
The solution to this test is a mixture containing shock, rarefaction and contact discontinuity. We carried out the computation up to $ t=0.4 $. The numerical results are given in Fig~\ref{1D BL-1}. We can see that our results are of good quality with such a low resolution grid.

\subsection{1D Euler equations}		
In this subsection, we present the numerical results of some benchmark shock-tube tests for 1D Euler equation to verify our new scheme for resolving the shock wave, contact discontinuity and rarefaction wave in compressible gas flows

\begin{figure} [hp]
	\centering
	\includegraphics[width=3in]{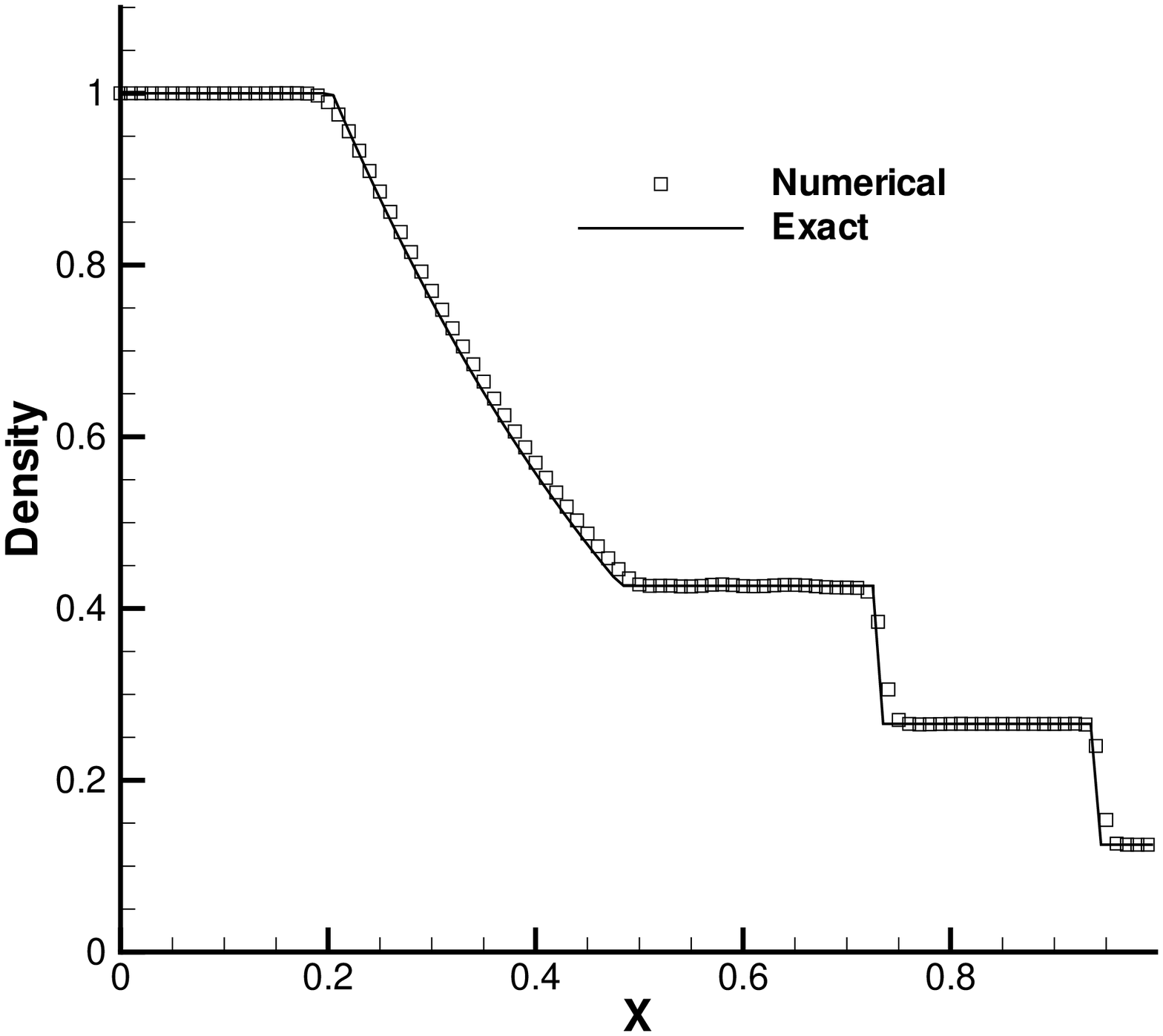}
	\caption{Numerical results of Sod's problem at $ t=0.25 $ with $ 100 $ cells.}
	\label{Sod's problem}
\end{figure}

\textbf{Sod's problem.} The initial condition is given by
\begin{equation}
(\rho_{0},v_{0},p_{0})=
\left\lbrace 
\begin{aligned}
&\hspace{0pt}(1,0,1), \quad \quad\quad 0\leqslant x\leqslant 0.5,\\
&\hspace{0pt}(0.125,0,0.1), \quad \rm{otherwise}.\\
\end{aligned}
\right.
\end{equation}
The mesh number is 100 in this test. The results presented in Fig.~\ref{Sod's problem} show that BVD-WENO-THINC scheme gives the accurate solution for contact discontinuity, shock and rarefaction wave. The contact discontinuity is resolved sharply.

\begin{figure}[hp]
	\centering
	\includegraphics[width=3in]{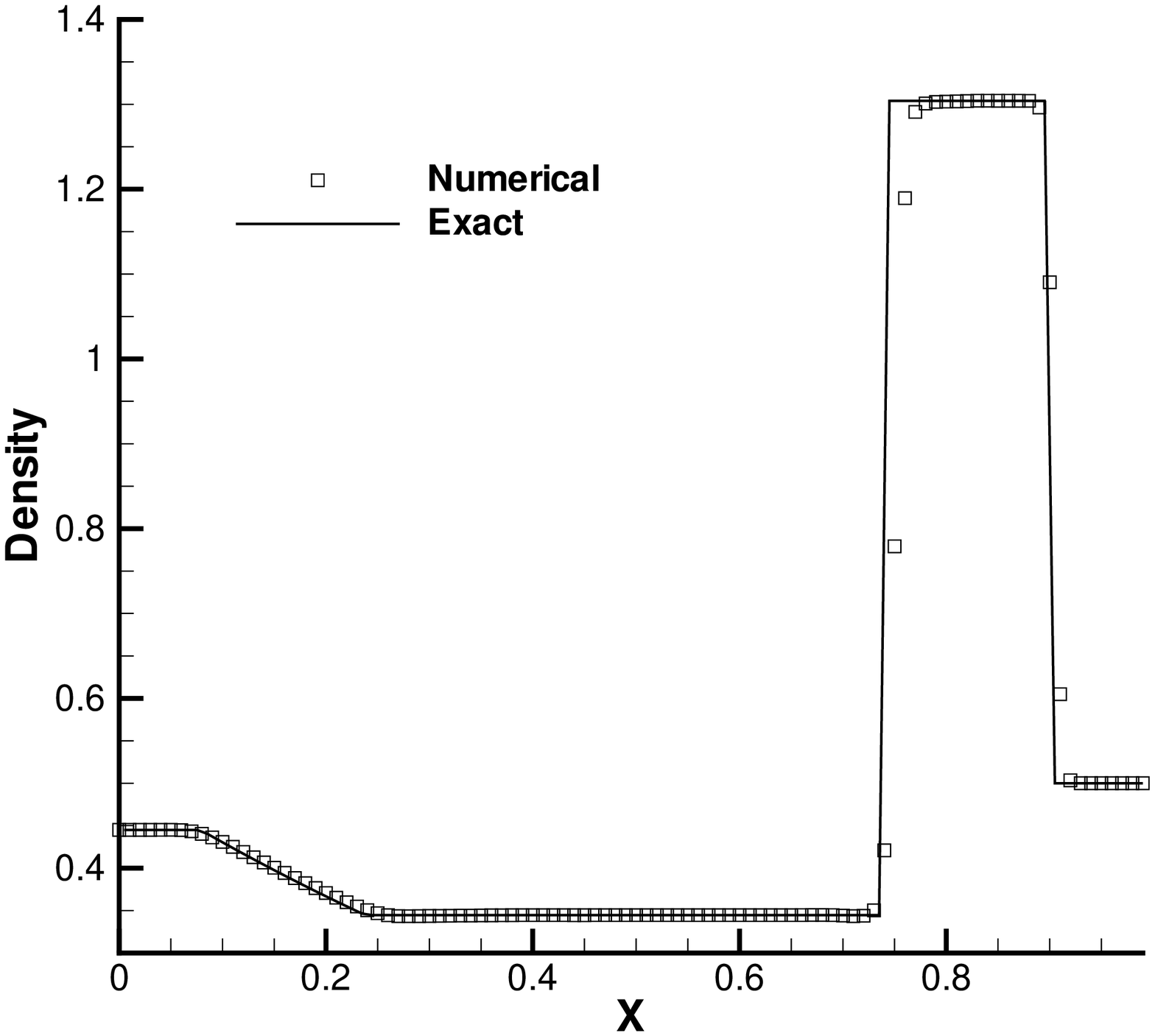}
	\caption{Numerical results of Lax's problem at $ t=0.16 $ with $ 100 $ cells.}
	\label{Lax's problem}
\end{figure}

\textbf{Lax's problem.} This is another widely used benchmark test with the initial condition given by
\begin{equation}
(\rho_{0},v_{0},p_{0})=
\left\lbrace 
\begin{aligned}
&\hspace{0pt}(0.445,0.698,3.528), \quad  0\leqslant x\leqslant 0.5,\\
&\hspace{0pt}(0.5,0,0.571), \quad\quad\quad\quad \rm{otherwise}.\\
\end{aligned}
\right.
\end{equation}
We show the numerical results at $ t=0.16 $ in Fig.~\ref{Lax's problem}. It can be seen again that our scheme can reproduce sharp contact discontinuity and shock wave, while maintaining the high accuracy for smooth profile.

\begin{figure}[hp]
		\centering
		\includegraphics[width=8cm]{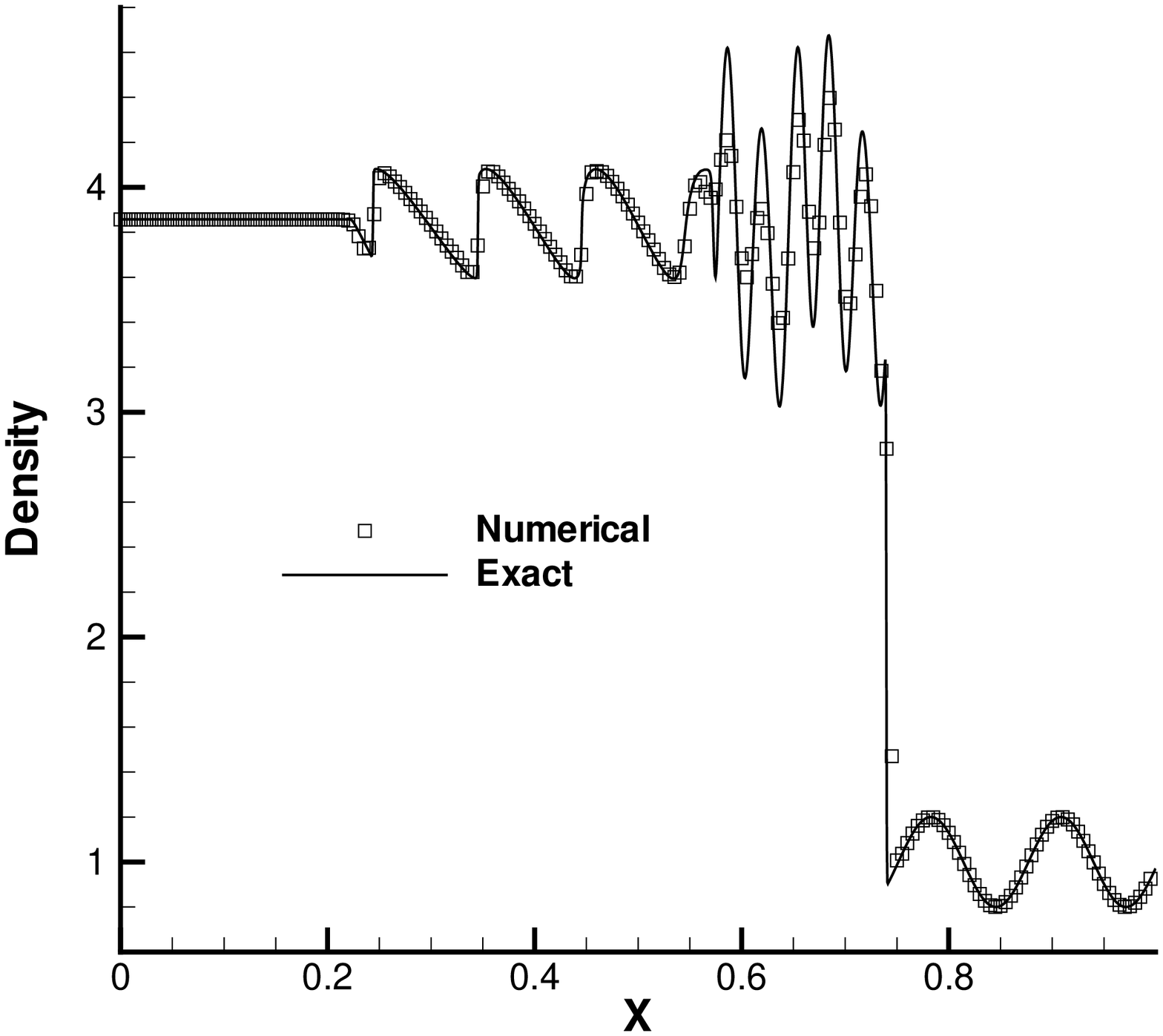} 
		\caption{Numerical results of shock-turbulence interaction at $ t=0.18 $ with $ 200 $ cells.}
		\label{shock-turbulence interaction}
	\end{figure}

\textbf{Shock-turbulence interaction.} The initial condition of this test is as follows
\begin{equation}
(\rho_{0},v_{0},p_{0})=
\left\lbrace 
\begin{aligned}
&\hspace{0pt}(3.857143,2.629369,10.333333), \quad \  \ \rm{when}\ x< 0.1,\\
&\hspace{0pt}(1.0+0.2\sin(50x-25),0,1.0), \quad\quad  \ \ \rm{when}\ x\geqslant 0.1.\\
\end{aligned}
\right.
\end{equation}
The computational domain is $ [0,1] $ and we perform the calculation up to $ t=0.18 $ on a mesh of $ 200 $ cells. In this test, the density perturbation can be easily smeared out when the numerical dissipation is significant. The numerical results shown in Fig~\ref{shock-turbulence interaction} reveal that our scheme can effectively reduce the numerical dissipation for the density perturbation and remove the spurious oscillation for shock wave.   

\begin{figure}[hp]
		\centering
		\includegraphics[width=8cm]{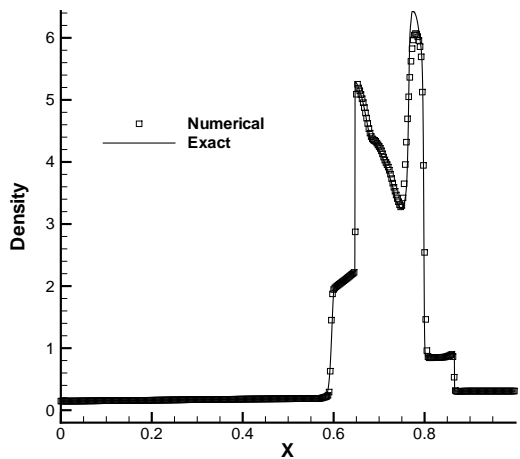} 
		\caption{Numerical results of two interacting blast waves at $ t=0.038 $ with $ 400 $ cells.}
		\label{Two interacting blast waves}
	\end{figure}

\textbf{Two interacting blast waves.} This is a 1D benchmark test including more complex structures. Initially we set the following condition for state variables
	\begin{equation}
	(\rho_{0},v_{0},p_{0})=
	\left\lbrace 
	\begin{aligned}
	&\hspace{0pt}(1,0,1000), \quad   0\leqslant x\leqslant 0.1,\\
	&\hspace{0pt}(1,0,0.01), \quad    0.1 < x< 0.9,\\
	&\hspace{0pt}(1,0,100), \quad  \ \ \ \rm{otherwise}.\\
	\end{aligned}
	\right.
	\end{equation}
The reflection boundary is set to the left and right ends of computational domain. This test includes the multiple interaction of strong shock waves and rarefactions. The common headache to nearly all existing shock-capturing schemes is the overly smeared density discontinuity in the numerical solution. The computed density is given in Fig.~\ref{Two interacting blast waves}. To our best knowledge, our result of this test is among the best ever reported in literature. Especially, the resolution of the left contact discontinuity is resolved within less than four cells, which is much superior to any other existing shock capturing scheme without artificial compression treatment. 

\subsection{2D Euler equations}
We solved two benchmark tests\cite{woodward1984} for 2D Euler equations.  

\begin{figure}[hp]
	\centering
	\subfigure[]{\label{DoubleMach-1}
		\begin{minipage}[b]{1.0\textwidth}
			\includegraphics[width=\textwidth]{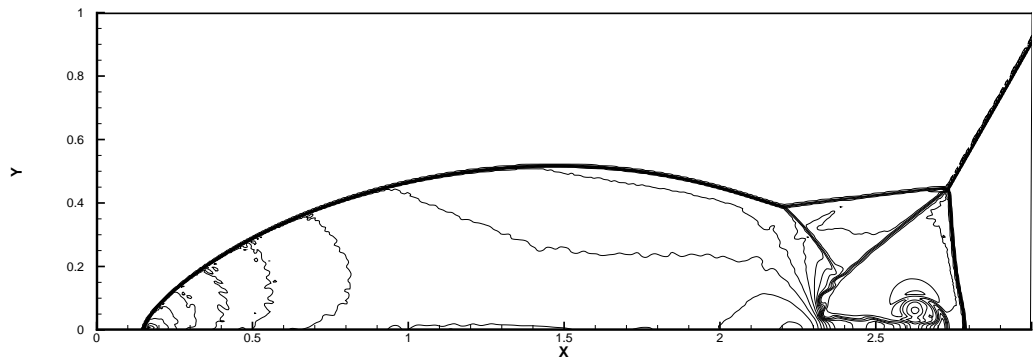} 
		\end{minipage}
	}
	\subfigure[]{\label{DoubleMach-2}
		\begin{minipage}[b]{1.0\textwidth}
			\includegraphics[width=\textwidth]{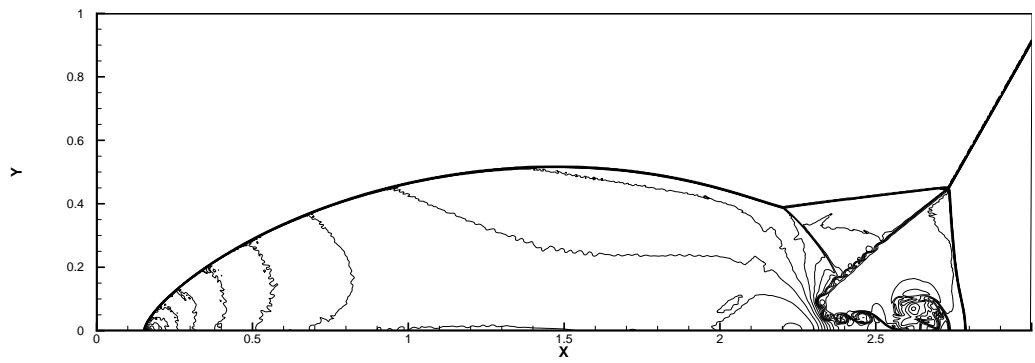} 
		\end{minipage}
	}
	\caption{Numerical results of double Mach reflection at $ t=0.2 $ with $200\times 640$ cells(a), $400\times 1280$ cells(b). }
	\label{DoubleMach}
\end{figure}

\begin{figure}[hp]
\centering
\subfigure[]{\label{DoubleMach-blowup-1}
	\begin{minipage}[b]{0.5\linewidth}
		\includegraphics[width=2.5in]{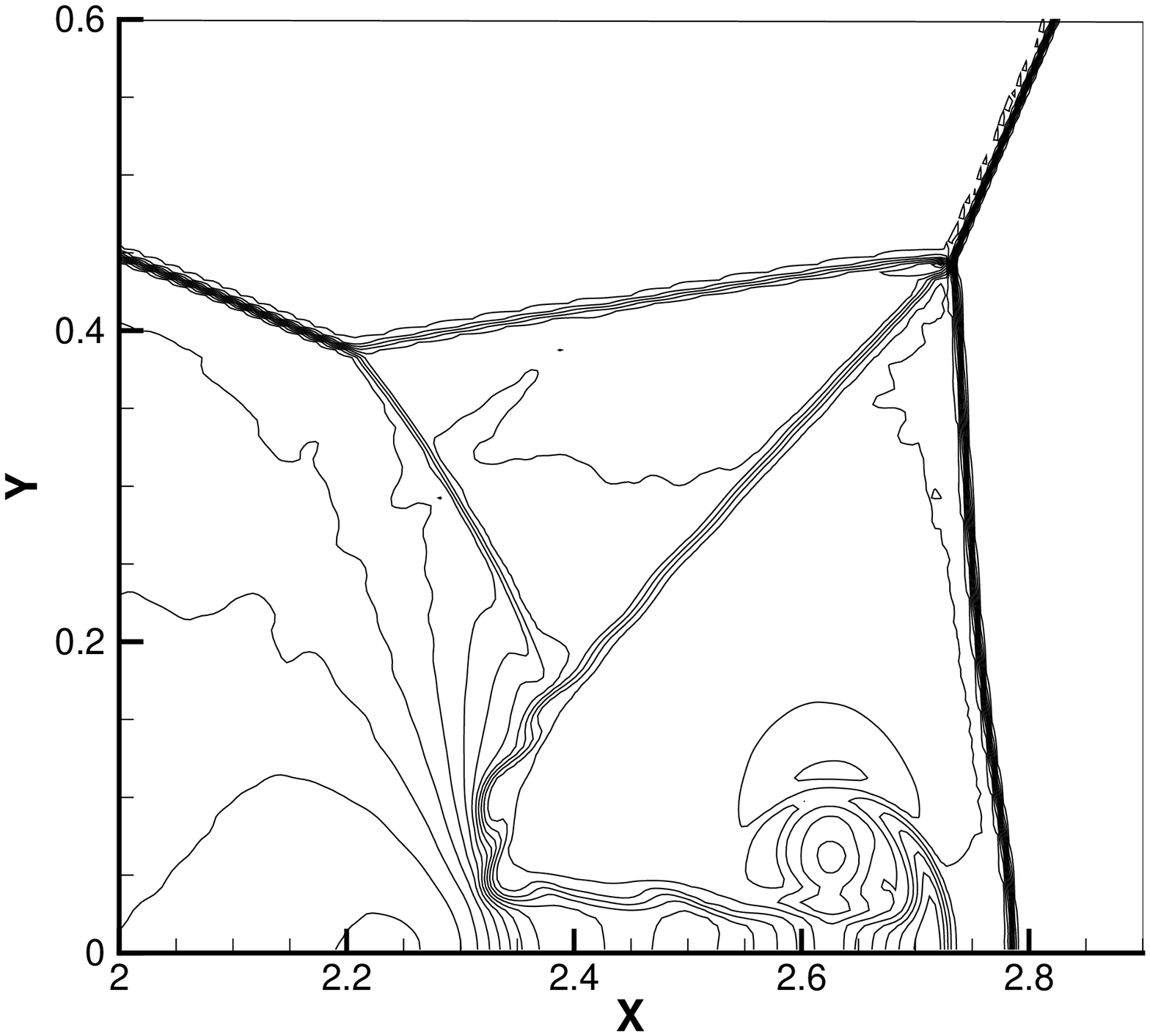} 
	\end{minipage}
}
\subfigure[]{\label{DoubleMach-blowup-2}
	\begin{minipage}[b]{0.3\linewidth}
		\includegraphics[width=2.5in]{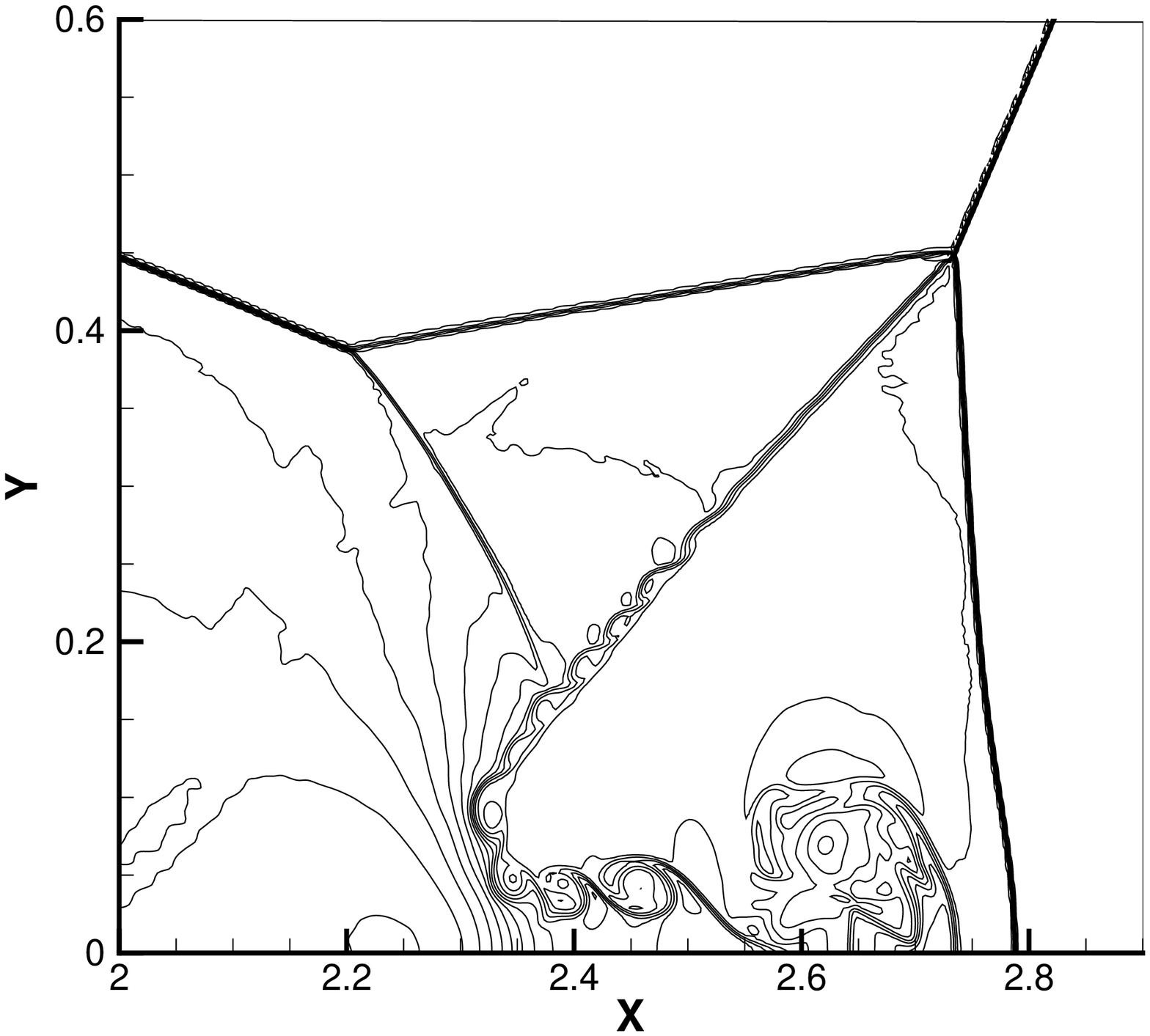} 
	\end{minipage}
}
\caption{Numerical results of double Mach reflection at $ t=0.2 $ with $200\times 640$ cells (a), $400\times 1280$ cells (b). }
\label{DoubleMach-blowup}
\end{figure}

\textbf{Double Mach Reflection.}
This benchmark contains both shock wave and contact discontinuity, and is usually used to verify numerical schemes in capturing strong shocks and vortical structures, which requires a numerical scheme to be optimized in terms of suppressing both numerical oscillation and dissipation.
 We computed the solution up to $ t=0.2 $. The numerical result computed on a mesh of $ 200\times640$ is shown in Figs \ref{DoubleMach-1} and \ref{DoubleMach-2} with part of the region zoomed in in Fig~\ref{DoubleMach-blowup-1} and \ref{DoubleMach-blowup-2} respectively. The numerical results are comparable to those obtained using discontinuous Galerkin (DG) method with even more degrees of freedom (DOFs)\cite{qiu2005,moe2015}.


\textbf{Mach 3 step tunnel.}
In this benchmark test, a right-moving Mach 3 flow in a tunnel with a forward step and reflective boundaries generates complex flows due to the interactions among shocks, density discontinuities and vortices. 

\begin{figure}[hp] 
	\centering
	\includegraphics[width=\textwidth]{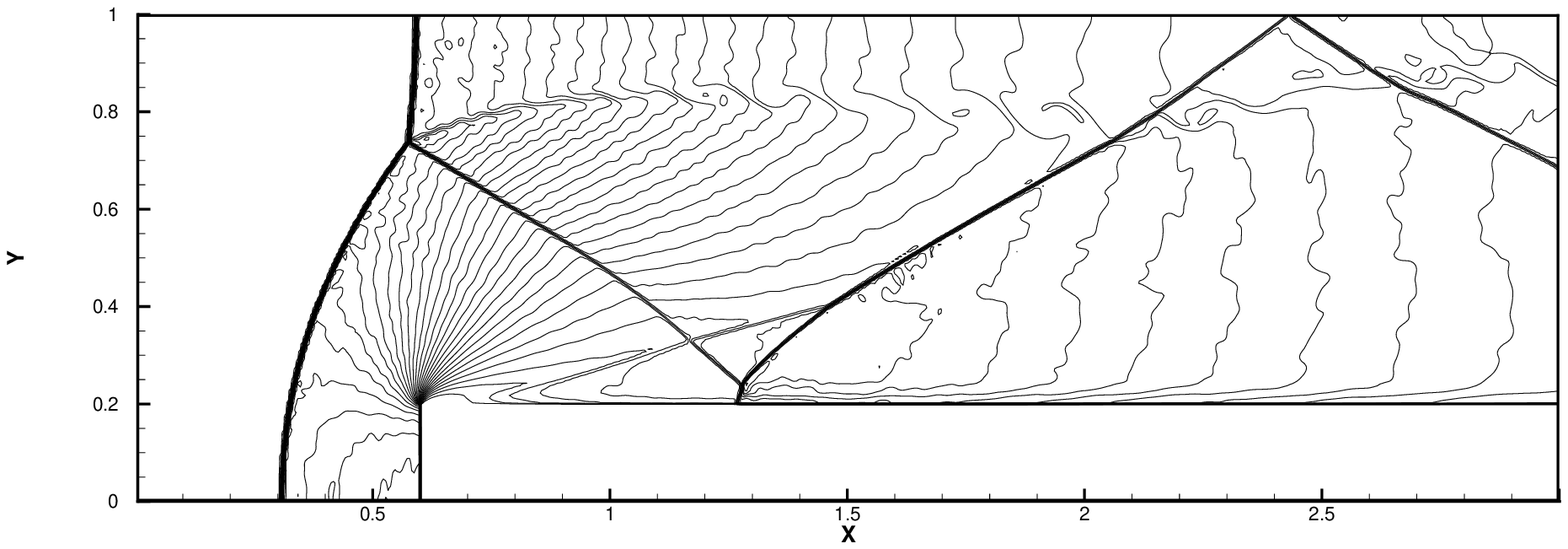} 
	\caption{Numerical results of Mach 3 step test at $ t=0.2 $ with $200\times 600$ cells. }
	 \label{Mach3-1}
\end{figure}

We show the BVD-WENO-THINC results at $T=4.0 $ on a $ 200\times 600 $ mesh in Fig~\ref{Mach3-1}. It can be seen again that the present scheme is able to give numerical solutions of competitive quality in comparison with DG models that use even more DOFs. 

\subsection{Summary}	

We have proposed a new guideline to construct high-resolution Godunov type schemes to resolve both smooth and discontinuous solutions. The basic idea, so-called  BVD (boundary variation diminishing), is to reconstruct the solution functions so that the jumps at cell boundaries are minimized, which effectively reduce the numerical dissipation in the resulting schemes. The  BVD reconstruction automatically realizes the highest possible polynomial interpolation for smooth profile, whilst prefers other forms of reconstructions in the presence of discontinuities. As a result, the  BVD reconstruction mitigates the questionable premiss that discontinuities only appear at cell interfaces in the current paradigm of high-order Godunov schemes.

We have implemented the BVD algorithm with two building block schemes, i.e. the WENO-Z scheme and the THINC scheme. The BVD   algorithm provide a reliable switching mechanism to reconstruct the solution function for both smooth profile and discontinuity.    
Excellent numerical results have been obtained for both scalar the Euler conservation laws, which show a substantial improvement in comparison with existing methods. 

Not limited to WENO-Z and THINC schemes, the BVD concept as a general platform of more profound impact can be used with other candidate reconstructions to further explore high-fidelity schemes for capturing both smooth and discontinuous solutions.


\end{document}